# Global Pandemic's Influence on Cyber Security and Cyber Crimes

Somya Khatri[1]    Aswani Kumar Cherukuri[1*]    Firuz Kamalov[2]

[1]School of Information Technology and Engineering, Vellore Institute of Technology, Vellore
Faculty of Engineering, Canadian University Dubai, UAE

* cherukuri@acm.org

**Abstract.** COVID-19 has caused widespread damage across many areas of life and has made humans more dependent on the internet and technology making us realize the importance of secure remote working environments. While social separation is encouraged during moments of lockdown, online infrastructure has become the central focus for communication, commerce, working, and learning, creating a new challenge and trend for companies to adopt new methods and operating models. The cases of cyber-attacks increased, and fraudsters and cybercriminals took use of this to intensify their illegal activities by taking advantage of remote workers' vulnerabilities and the public's interest in information about the coronavirus. This paper examines the different types of security threats and cyber-crimes that people faced in the pandemic time and the need for a safe and secure cyber infrastructure. This paper attempts to analyze the security implications of the issues.

## 1      Introduction

The COVID-19 pandemic has ushered in a sea of change that has affected nearly every aspect of our life. Since March 2020, global internet usage has increased by 50-70%. Cybersecurity has not been spared from these developments, and it now faces a whole new set of challenges. The transition to remote employment, for example, has opened the door to a variety of attack vectors. Furthermore, the pandemic's fear, confusion, and disinformation have provided chances for cybercriminals to conduct phishing scams, ransomware assaults, and other malicious activities. We can frequently see patterns forming in cybersecurity by looking at current developments. However, because things change so rapidly, even research from 2019 doesn't accurately portray the challenges we face now. Fortunately, much effort is being done to assess the current situation so that we may better prepare for the post-pandemic environment.

   Our dependency on the cyber infrastructure has increased and therefore the threats have also multiplied. With the introduction of hybrid modes of working, it becomes extremely important for us to take measures to mitigate the effects of cyber security threats and attacks and to increase the security of our cyber infrastructure.

This article examines the different types of security threats and cyber-crimes that people faced in the pandemic time and what were its security implications. The motivation behind this work is to highlight the need for a safe and secure cyber infrastructure for future.

We have first provided the background for the study which talks about the effect on organizations, security policies, then there is detailed literature survey done on the existing work in the aspects of identification of threats, their mitigation and the available cyber infrastructure. It is followed by a summary of key security issues related to COVID-19 cybersecurity and how they escalated during the Covid-19 pandemic. We have then analysed the security implications these issues had. It is then followed by the conclusion of the study done.

## 2     Background

Cyberattacks have increased as a result of growth of communications and the shift to the digital mode. This has led to a number of additional dangers and vulnerabilities for information and systems. The security of organizations is at risk of being breached. For breaches that occur at both physical and digital access points, they need to have continuous surveillance and risk assessments in order to keep an eye on the situation and take appropriate action to prevent and mitigate the ill effects.

For many organizations, the IT department is under a lot of pressure due to this transition to remote employment, while ensuring the security of the information systems. Collaboration tools fulfill the requirement of keeping staff in synchronization, at the same time it makes the system more vulnerable to critical information of the organization being hacked as it is now stored in less secure remote locations.

Applying company security policies and controls to remote employees is difficult as they have limited scalability and can take a long time to set up. Business continuity plans (BCPs) and incident response plans (IRPs) are almost always inadequate or nonexistent. Security officials had never before attempted a BCP attack on this scale. The increased digital footprint and traffic are being used by cybercriminals to scan systems for weaknesses or steal money. They are well aware that numerous businesses and workers have exposed themselves to hacking. For instance, phishing emails with attachments that download malware, crash computers, or steal sensitive data are being sent out. They are themed on COVID-19. Attackers build fictitious websites or hijack weak ones to house malware. The websites draw visitors, who then download malicious malware onto their machines as a result. Some fraudulent websites also solicit payments from paid workers through email connections. Numerous Covid-19 patient count-status programs and URLs have been identified to include viruses and malware that steals identities.

## 3   Literature Survey

The scientific community has begun to pay attention to the problem of COVID-19 cyberattacks. There has been a lot of research work that has been done in the cyber security field which we can relate to the covid-19 scenario. We have classified the literature survey done into three categories broadly: Table 1 for identification of threats, Table 2 their mitigation or prevention and Table 3 for research on the cyber security infrastructure available.

**Table 1.** Identification of threats

| Reference | Methodology/Approach | Remarks/Observations |
|---|---|---|
| [1] | Pattern identification and detection, penetration testing, and cyber-crime mapping | Automating schedule assignments to improve security processes, allowing security examiners to focus on data that requires significant programming abilities. |
| [2] | 5 types of attacks were examined and related to the Cyber Kill Chain model, as well as protection techniques against these attacks. | The exploration can be adapted to future X-themed cyberattacks exploiting future events. |
| [3] | Exploratory data analysis performed on 155k shared threats including 10k cyber threats related to COVID-19. | Should be investigated further to determine if a prediction model can be utilized to corroborate the linkages and patterns. |
| [4] | Fuzzy logic and data mining -based intelligence system for detecting Covid-19 themed malicious phishing attacks. | Systems must be trained with high-quality inputs that represent desirable. It is far less predictable in real-world circumstances. |

**Table 2.** Mitigation

| Reference | Methodology/Approach | Remarks/Observations |
|---|---|---|
| [5] | Approaches to overcome cyber threats(anti-virus software, firewalls, updated OS) | Skilled hackers can exploit weakness in network configurations and bypass it. |
| [6] | Neural Network to collect data about phishing attacks | New IoT layered models with privacy and security components and layer recognition. This may be used in con- |

| Reference | | |
|---|---|---|
| | | junction with security measures to mitigate cybersecurity threats at each layer: IoT, edge, and cloud. |
| [7] | Examines high probability data security threats and mitigations | Gain insights into common network security concerns. Not enough data sources. |
| [8] | SDN-based networks to flexibly solve security and management problems | SDN becomes into a single point of failure, making it a valuable target for attackers. |
| [9] | examines the most typical wireless network attacks and the security breaches that have been reported throughout the covid-19 period | Implementing the correct security measures for the wireless network may be expensive. |

**Table 3.** Cyber Infrastructure

| Reference | Methodology/Approach | Remarks/Observations |
|---|---|---|
| [10] | The report examines the consequences and makes recommendations to improve the marine transportation system's security and resilience in the event of future disruptions. | Such an analytical advantage might be used by MTS sub-sectors to identify how to best prioritize security enhancements, encouraging improved resilience in the industry as a whole. |
| [11] | Describes company's evaluation of critical threats in the on-site working paradigm versus the remote working paradigm | A ubiquitous secure technological system model that protects the company's network architecture as well as access to secret data will need to be built. |
| [12] | Examine a university's current remote access(VPN) security vulnerabilities | Any university can use it as a broad security guideline to assess the security of its remote access and Internet border system. High cost for installing these devices and software for everyone on campus |
| [13] | Development of a secure online learning system, which utilizes edge computing and privacy mechanisms such as trust evaluation from direct and indirect observations. | Used the Bayesian inference technique and the Dempster Shafer theory |
| [14] | Comprehensive analysis of 70 CT apps | 80% handled sensitive data without adequate protection and weak cryptographic algorithms and embedded data trackers |
| [15] | Identify cybersecurity problems and present alternatives to | Prototype of a National Blockchain-based Public Data Registration System |

| improve IS in public organizations. | (SNRDP-DLT) to help better manage risk from natural and man-made catastrophes. |

## 4 Security Issues

Cybercrime is defined as any criminal activity that takes place through the use of a networked device or a network and a computer. Most of them involve stealing or damaging information to make money for the criminals, but there are also cases where they do so to harm people. Additionally, there are cybercrimes that disseminate viruses, illicit data, photographs, and other items via computers and networks.

Below is a summary of key security issues related to COVID-19 cybersecurity and how they escalated during the Covid-19 pandemic and what to expect in future.

### 4.1 Identity theft

As a result of the COVID-19 outbreak, we were obliged to go online. When we purchase online, seek employment, or deal with crucial affairs, we give up our personal information. Attackers have the ability to perform a variety of things to a victim's identity. They can take control of their accounts accreditation, open new money accounts, and steal their hard-earned money. The attackers' main goal is to obtain critical pieces of information about them in order to persuade a bank that they are the person. Clients should be careful not to provide too much personal information about themselves on social media and other platforms. The victim must also keep all financial information hidden. On any social networking platform, personal information should not be shared.

### 4.2 Hacking

Hacking is gaining access to the system structure without the owners or users consent. Interlopers are mostly software engineers with advanced understanding of computer systems who exploit this expertise for nefarious ends. Some snoopers do it casually to demonstrate their proficiency, which may range from executing secure transactions to changing computer programmes to complete projects that exceed the creator's expectations. During the current COVID-19 pandemic, the criminal-minded threat actors driving the bulk of cyberattacks have updated their attacking techniques. Since the epidemic began, fraudsters have been building new phishing tools, hacking methods, and experimenting with new attack paths in order to profit from the crisis and demonstrate their cyber expertise.

### 4.3 Denial of Service attack

An uncompromising endeavour in which the attackers refuse to provide services to customers who are entitled to them is a Dos attack. A Dos attack overloads the com-puter's asset by flooding it with requests that are more than the computer can handle, exhausting the computer's available transfer speed and overburdening its server. This causes the framework to fail, with the server temporarily failing or crashing altogether. The greatest notable rise in DDoS assaults was seen in March and April during the pandemic times. In comparison to the first quarter of 2019, the number of assaults has climbed by 542.46%, with a QoQ increase of 278.17%. DDoS assaults have traditionally been "off-season" in the first quarter. While this is an intriguing divergence, we feel that the current COVID-19 pandemic may be a key factor.[16]

### 4.4 Phishing

By impersonating an authorized user, private data of the user like the password, phone number may be extracted. It's a sort of social engineering that's frequently used in the form of mail hoaxing. The victims are persuaded to download malware from the internet by social ads, and the cybercriminals force them to give out personal information under false pretenses. It is also a common way to propagate malware in, by allowing victims to download a report or click on a link that would secretly install the malicious payload in assaults that may include trojan virus, ransomware, or other types of harmful and bothersome attacks.

According to a new study by F5 Labs, COVID-19 continues to contribute to the phishing and fraudulent activities of cybercriminals. It was shown that during the height of the outbreak, phishing incidents rose by 220% when compared to the annual average. The three primary goals of COVID-related phishing emails have been identified as credential collection, malware dissemination, and fraudulent donations to fictitious charity.

### 4.5 Digital Stalking

Type of cybercrime that includes online torture, in which the victim is harassed by a barrage of tweets and emails. Typically, cyber tormentors use social networks and websites to intimidate and instil terror in their victims. The cyber tormentors are well aware of the harm their actions have resulted in, and the goal is to make the victim bewildered and concerned about their safety.

Pandemic has increased our dependence on digital tools and increased the incidence of cyber harassment. Therefore, in order to increase the protection of online users and to stop digital abuse, legal protection and other measures must be urgently adopted to address this new reality. Civil communities can as well act as an important role in increasing public awareness, building the ability to detect and report the cases of cyber harassment, and finding suitable tools and services to mitigate impact. In addition to the efforts of the international community, government and policy makers, changes in the law will give men and women equal access to a secure digital environment.

### 4.6 Exploiting machine learning

When a programmer attempts to put rules on a machine learning algorithm, the system and framework are open to attack. In order to learn for themselves, machine learning algorithms require information from social or community-driven networks. Snoopers try to use surveys, web surfing, and other online activities to collect personally identifying information. Cyber snoopers interested in ML exploitation may employ malicious tests or backdoors to access both the framework and the data required for training.

## 5 Security Implications

### 5.1 Cyber Infrastructure

Cyberinfrastructure is made up of computational systems, data and information management, cutting-edge instruments, visualization environments, and people. These components are connected by software and cutting-edge networks to increase productivity and facilitate discoveries and breakthroughs in knowledge that would not otherwise be possible. With the advent of technology, Internet of Things (IoT) and home automation, many of our devices are getting connected and our dependence on them has increased and so have the possible threats and attacks. To mitigate the effects of these threats several new authentication mechanisms have been introduced for example password-less authentication, token based authentication etc.

### 5.2 Government's role

Our dependency on digital technology has considerably expanded as a result of the COVID-19 pandemic. As remote working has become crucial to our economy and medical treatment, people and organisations will only continue to become more dependent on digital technology. The risk of cyberattacks, however, increases with every new device, user, and business that connects to the internet. In order for society and economies to advance, governments must be able to deliver dependable and secure digital connectivity. A robust national cybersecurity ecosystem, a dedicated national cyber-security agency, a National Critical Infrastructure Protection programme, a national incident response and recovery strategy, and clearly defined laws pertaining to all cybercrimes are the main elements of an extensive national cybersecurity plan that make up effective national cybersecurity strategies.

### 5.3 Hybrid Environment

Even if the pandemic has been going on for two years, there are still insufficient resources available. This presents a challenge to malicious individuals looking for weaker organizations. However, the pandemic's uncontrolled actions have also created new possibilities for stronger security and economic continuity. Increase in remote employees and clients has irreversibly altered the corporate landscape. Even businesses that are still holding onto commercial real estate realize they must deal with remote workers or hybrid workplaces for the foreseeable future. Some businesses closed their physical offices permanently. As a result, companies have increased their reliance on traditional network security measures like VPNs and firewalls installed on workplace mobile devices and transferred more apps into the cloud. Many businesses are implementing systems to monitor and control DNS, DHCP, and IP traffic moving into and out of servers for employees using their own equipment. Emerging hybrid workforce reality is producing more problems with assaults, ransomware, and data leaking. The cloud, employee-owned endpoints, and WiFi access points were the usual starting locations for attacks. Phishing was a popular method of gaining unauthorized access to take over credentials and steal or lock down data files.

The use of Secure Access Service Edge (SASE) frameworks is growing. To protect sensitive information from the hazards associated with remote employees using a mix of personal and company devices, several organisations added smartphones to their equipment fleets, giving corporate-owned mobile devices to slightly more than half of respondents. A similar number of people used virtual private networks to encrypt internet traffic, especially when a remote worker was using an unprotected or unsecured wireless network. Along with these and other procedures were developed to provide personnel with safe tools to perform their responsibilities under difficult circumstances without disturbing the organization. Some of the most popular security procedures are brokers for secure cloud access (CASB) data loss prevention through data protection Endpoint DNS safety detection and response monitoring, detecting, and reacting to network activity for network protection, de-provisioning and provisioning trustworthy web gateways, VPNs, and other access control devices in a secure manner.

### 5.4 Cyber hygiene and awareness

Cyber hygiene is a set of consistent practises used by organisations and people to safeguard users, devices, networks, and data. A company can lower the risk of operational disruptions, data compromise, and data loss by strengthening its overall security posture and using excellent cyber hygiene. The entire effectiveness of a company's cybersecurity programme determines how well-prepared it is to counter both current and future threats. This is referred to as an organization's security posture. Use proper cyber hygiene to gain the best cybersecurity. Threats to cyber hygiene include user buy-in, predictability, and the breadth and complexity of IT systems. Users need to be aware of the best cyber hygiene practices that is to take regular backups, educate themselves about how to prevent common attacks, use encryption to protect sensitive data, make sure firewalls and routers are properly set, maintain good password hygiene, use MFA (Multi Factor authentication), patch management, online discretion and security software.

## 6     Conclusions

The study made an effort to concentrate on the existing cyber security threats and weaknesses in relation to the COVID-19 worldwide pandemic. The greatest internet use ever occurred as a result of this pandemic. Everyone was required to utilize the internet to maintain their communication, commerce, and education across all age groups and walks of life. As so many people use the internet, cybercriminals have taken advantage of this to make money. The threats of cyber security have substantially increased during this pandemic due to a lack of understanding in this area.

It is necessary to have a fundamental grasp of all types of potential cybersecurity threats and cyberattacks. To avoid crucial data getting into the hands of cybercriminals, it is imperative that all workers have a basic awareness of cybersecurity. The corona virus pandemic is only the start. These viruses may become more prevalent in the coming generations. It is now appropriate to consider the future. To lessen and ameliorate the challenges brought on by the COVID-19 epidemic, we must all be able to learn from it and get ready for the future.